\documentclass[12pt]{article}

\usepackage{graphicx}

\newenvironment{sciabstract}{%
\begin{quote} \bf}
{\end{quote}}

\newcounter{lastnote}
\newenvironment{scilastnote}{%
\setcounter{lastnote}{\value{enumiv}}%
\addtocounter{lastnote}{+1}%
\begin{list}%
{\arabic{lastnote}.}
{\setlength{\leftmargin}{.22in}}
{\setlength{\labelsep}{.5em}}}
{\end{list}}

\title{Quantum Gas of Deeply Bound Ground State Molecules}

\author
{Johann G. Danzl$^1$, Elmar Haller$^1$, Mattias Gustavsson$^1$, Manfred J. Mark$^1$, \\
Russell Hart$^1$, Nadia Bouloufa$^2$, Olivier Dulieu$^2$, Helmut Ritsch$^3$, \\
Hanns-Christoph N\"agerl$^1$\\
\\
\normalsize{$^1$Institut f\"ur Experimentalphysik und
Zentrum f\"{u}r Quantenphysik,}\\
\normalsize{Universit\"at Innsbruck,}\\
\normalsize{Technikerstra{\ss}e 25, 6020 Innsbruck, Austria}\\
\\
\normalsize{$^2$Laboratoire Aim\'e Cotton,}\\
\normalsize{CNRS, Universit\'e Paris-Sud}\\
\normalsize{B\^{a}t. 505, 91405 Orsay Cedex, France}\\
\\
\normalsize{$^3$Institut f\"ur Theoretische Physik und
Zentrum f\"{u}r Quantenphysik,}\\
\normalsize{Universit\"at Innsbruck,}\\
\normalsize{Technikerstra{\ss}e 25, 6020 Innsbruck, Austria}\\
\\
}

\date{}

\begin{document}

\baselineskip24pt

\maketitle


\begin{sciabstract}
We create an ultracold dense quantum gas of ground state molecules bound by more than 1000 wavenumbers by stimulated two-photon transfer of molecules associated on a Feshbach resonance from a Bose-Einstein condensate of cesium atoms. The transfer efficiency exceeds 80\%. In the process, the initial loose, long-range electrostatic bond of the Feshbach molecule is coherently transformed into a tight chemical bond. We demonstrate coherence of the transfer in a Ramsey-type experiment and show that the molecular sample is not heated during the transfer. Our results show that the preparation of a quantum gas of molecules in arbitrary rovibrational states is possible and that the creation of a Bose-Einstein condensate of molecules in their rovibronic ground state is within reach.
\end{sciabstract}




Ultracold samples of molecules are ideally suited for fundamental studies in physics and chemistry, ranging from few-body collisional physics \cite{Chin2005,Kraemer2006,Staanum2006,Zahzam2006}, ultracold chemistry \cite{Krems2005}, high resolution spectroscopy \cite{Zelevinsky2008,DeMille2008}, to quantum gases and molecular Bose-Einstein condensation \cite{Fermi2008} and quantum processing \cite{DeMille2002}. For many of the proposed experiments full control over the molecular wave function in specific deeply bound rovibrational states is needed. High densities are required for molecular quantum gas studies. Only in the rovibronic ground state, i.e. the lowest energy level of the electronic ground state, is collisional stability assured. However, direct molecular cooling towards high phase space densities seems yet out of reach \cite{Doyle2004}, whereas techniques like Feshbach association \cite{Koehler2006} and photoassociation \cite{Jones2006} either only produce molecules in weakly bound rovibrational levels or suffer from low production rates and low state selectivity. In order to produce a quantum gas of molecules in their absolute ground state, Jaksch {\em et al.} \cite{Jaksch2002} proposed a scheme for homonuclear alkali molecules in which the technique of stimulated two-photon transfer is repeatedly applied to molecules associated from a high-density sample of ultracold atoms. The initially very loosely bound molecules are transferred in successive steps to the rovibrational ground state of the singlet $X^1\Sigma_g^+$ molecular potential. The advantage of this scheme is that it is fully coherent, not relying on spontaneous processes, and that it involves only a very small number of intermediate levels. It promises that a ground state binding energy of typically 0.5 eV can be carried away without heating the molecular sample. It essentially preserves phase space density, allowing the molecular sample to inherit the high initial phase space density from the atomic sample. However, to realize this scheme, several challenges have to be met. First, there is a large difference in internuclear separation that has to be bridged, i.e. the overlap between the radial wave function of the least bound molecules with the radial wave functions of deeply bound molecular levels is extremely low, potentially leading to prohibitively low transition rates for the two-photon transitions. Second, the scheme requires the identification of suitable intermediate molecular levels while strictly avoiding parasitic excitations. Third, a large difference in binding energy has to be overcome. On a more technical side, the lasers driving the two-photon transitions at widely different wavelengths need to have extremely low relative short term phase jitter and high long term frequency stability to allow for coherence and reproducibility. In important experiments, Winkler {\it et al.} \cite{Winkler2007} and recently Ospelkaus {\it et al.} \cite{Ospelkaus2008} demonstrated highly efficient two-photon transfer into lower lying molecular levels starting from weakly bound dimer molecules, which were associated from ultracold atoms on a Feshbach resonance \cite{Koehler2006}. However, the transferred molecules are still weakly bound. Their binding energy is less than $10^{-4}$ of the binding energy of the rovibrational ground state, and wave function overlap with this state is still negligible.

Here we demonstrate {\em the} crucial step towards full control of the molecular wave function and towards the formation of a Bose-Einstein condensate (BEC) of molecules in their rovibronic ground state by linking weakly bound molecular states with deeply bound rovibrational states. We coherently transfer an ultracold quantum gas of weakly bound cesium Feshbach molecules to the rovibrational level $|\nu\!=\!73, J\!=\!2\!\!>$ of the singlet $X^1\Sigma_g^+$ potential, bound by $ 1061 $ cm$^{-1}$, or $ h\!\times\!31.81 $ THz, more than one fourth of the binding energy of the rovibrational ground state. As important ingredients, we overcome low wave function overlap by using a suitable intermediate excited molecular state while avoiding excitation into loss channels, and we reference the transfer lasers to a frequency comb, allowing us to flexibly bridge binding energy differences of more than 1000 cm$^{-1}$ while assuring relative coherence.

Fig. 1 shows the energy of the relevant molecular and atomic states. Our experiment starts with a cigar-shaped BEC of cesium atoms in the lowest hyperfine sublevel $F\!=\!3, m_F\!=\!3$ in an optical dipole trap. For BEC production, we essentially follow the procedure detailed in Ref.~\cite{Weber2003}. For Feshbach molecule production out of the BEC, we ramp up the offset magnetic field from the initial value of $2.1$ mT to about $5.0$ mT in $ 10 $ ms. We then ramp down, sweeping across a d-wave Feshbach resonance at 4.8 mT after about $1$ ms as shown in Fig. 1B \cite{Herbig2003,Mark2007}. Our procedure \cite{Herbig2003} gives an ultracold and dense sample of up to 11000 molecules every 10 s. Upon lowering the magnetic field, the molecules are transferred from the initial state $|d\!\!> $ to a still weakly bound s-wave molecular state $|s\!\!> $ of the lowest hyperfine channel $(F_1\!=\!3, F_2\!=\!3)$ via an avoided crossing \cite{Mark2007}. The index $i\!=\!1,2$ denotes the $i$-th atom. Upon further lowering the magnetic field to about 2.2 mT, the molecules enter into a closed channel s-wave molecular state $ |a\!\!> $ via a second, broad avoided crossing \cite{Mark2007}. This state belongs to the uppermost hyperfine channel $(F_1\!=\!4, F_2\!=\!4)$ and thus has an effective binding energy of more than $ 2\times h \nu_\mathrm{Cs} $. Here $ h $ is Planck's constant and $\nu_\mathrm{Cs}\!\approx\!9.19$ GHz is the Cs clock frequency. Similar to $|s\!\!> $ this state is a mixture of the $X^1\Sigma_g^+$ ground state and the lowest triplet $a^3\Sigma_u^+$ state, coupled by hyperfine interaction, and it has zero rotational angular momentum. At a field of 1.9 mT, it has a binding energy of $ 5 $ MHz$\times h$ with respect to the $F\!=\!3, m_F\!=\!3$ two-atom asymptote \cite{Mark2007}. As one might expect, we find that optical transition rates as measured below are improved when using this effectively more deeply bound state as the initial state for two-photon transfer instead of state $|s\!\!> $. We shut off the trap and perform all subsequent experiments in free flight to reduce the particle density in order to avoid collisional processes. We detect molecules in $|a\!\!>$ via states $|s\!\!> $ and $|d\!\!> $ by reversing the magnetic field ramp, dissociating them on the Feshbach resonance at 4.8 mT, and imaging the resulting atoms \cite{Herbig2003}.

Efficient two-photon transfer via the stimulated Raman adiabatic passage (STIRAP) technique \cite{Bergmann1998,Winkler2007} relies on a suitable choice for the excited state $|e\!\!>$. In our case this state must have singlet character so that it can be used as a transfer state to deeply bound levels of the $X^1\Sigma_g^+$ potential. In general, it must be well separated from other states, which otherwise could be off-resonantly excited. It should thus be situated far to the red of the excited S$_\frac{1}{2}$+P$_\frac{1}{2}$ potential asymptote to avoid the high density of excited molecular states near that asymptote. We have performed optical loss spectroscopy starting from state $|a\!\!>$ in the wavelength range from 1120 to 1130 nm, about 2300 cm$^{-1}$ to the red of the cesium D$_1$ line. For this we recorded the number of remaining molecules in $|a\!\!>$ as a function of excitation wavelength and found two progressions of lines, which we assign to the potential curves of the mixed $(A^1\Sigma_u^+ - b^3\Pi_u) \ 0_u^+$ excited states and to the $(1)^3\Sigma_g^+$ excited state, respectively. For the present experiments, we choose for $|e\!\!>$ a level of the $ 0_u^+ $ progression which is $8879.63(1)$ cm$^{-1}$ above the $F\!=\!3, m_F\!=\!3$ two-atom asymptote, corresponding to a transition wavelength of $ 1126.173(1) $ nm, see Fig.~1A. We measure all wavelengths on a home-built wavemeter. We identify this previously unknown level as the 225th one of the $ 0_u^+ $ system, with about two units of uncertainty.

The ground state level $|g\!\!>$ with vibrational quantum number $\nu\!=\!73$ is well known from conventional molecular spectroscopy \cite{Weickenmeier1985,Amiot2002}. However, its binding energy, as well as the binding energy of all vibrational levels, has only been known with an uncertainty of about $\pm 0.45$ cm$^{-1}$ prior to the present experiments \cite{Amiot2002}. We search for $|g\!\!>$ by exciting the transitions from $|a\!\!>$ to $|e\!\!>$ with laser $L_1$ and from $|e\!\!>$ to $|g\!\!>$ with laser $L_2$ simultaneously \cite{Lasers}. The two light fields create a molecule-molecule dark state. The molecules initially in $|a\!\!>$ are lost unless the second laser $L_2$ is on two-photon resonance, provided that the Rabi frequency $\Omega_2$ on the second transition is equal to or greater than $\Omega_1$, the Rabi frequency on the first transition. For coherence, stability, and reproducibility, we lock both lasers to independent narrow-band optical resonators, which we reference to an optical frequency comb. The comb is not calibrated, but it allows precise differential frequency measurements and provides long-term stability needed for systematic line searches \cite{wavemeter}. We find the resonance condition with vibrational level $\nu\!=\!73$ at $1005.976(1)$ and $1005.982(1)$ nm, corresponding to rotational quantum numbers $J\!=\!0$ and $2$. Identification of $J$ is possible since the rotational energy splitting is well known. Fig.~2A and B show typical molecular dark resonances when we set $L_2$ on resonance and step the detuning $\Delta_1$ of $L_1$ near $1126.173$ nm. Fig. 2C shows a dark resonance involving $\nu\!=\!73, J\!=\!2$ using a different excited molecular state $|e'\!\!>$, which is excited with $L_1$ near $1123.104$ nm. Fig. 2D-F show dark resonances involving the neighboring vibrational levels $\nu\!=\!71$ and $\nu\!=\!72$. These $X^1\Sigma_g^+$-levels were easily found on the grounds of previous data \cite{Amiot2002}. We determine the binding energy of these levels with respect to the atomic $F_1\!=\!3, F_2\!=\!3$ asymptote at zero magnetic field to be $ 1060.9694(10), 1088.3101(10), 1115.9148(10) $ cm$^{-1}$ for $\nu\!=\!73,72,71$ with $J\!=\!0$, respectively. The binding energy of the rovibrational ground state $\nu\!=\!0$ is thus $3628.7053(10)$ cm$^{-1}$, which represents an improvement of 3 orders of magnitude compared to the previous determination \cite{Amiot2002}. Fitting the data for the dark resonances with a three-level model taking into account off-resonant excitations and laser line widths, we determine the molecular transition strengths as given by the normalized Rabi frequencies for the transitions $|a\!\!>$ to $|e\!\!>$ and $|e\!\!>$ to $|\nu\!=\!73,J\!=\!2\!\!>$ to be $\Omega_1\!=\!2\pi\!\times\!2$ kHz $ \sqrt{I/(\mathrm{mW/cm}^2)}$ and $\Omega_2\!=\!2\pi\!\times\!11$ kHz $ \sqrt{I/(\mathrm{mW/cm}^2)}$, respectively. A comparison with a typical atomic transition strength of $\Omega_a\!=\!2\pi\!\times\!5$ MHz $ \sqrt{I/(\mathrm{mW/cm}^2)}$ giving $ |\Omega_1/\Omega_a|^2 < 10^{-6} $ reflects the minuteness of the wave function overlap.

We are now in a position to carry out coherent transfer using the STIRAP technique. For $|g\!\!>$ we choose the vibrational level with $\nu\!=\!73, J\!=\!2$. This level will allow us to reach the rovibrational ground state $\nu\!=\!0, J\!=\!0$ with a second STIRAP step in view of the selection rule $\Delta J\!=\!0,\pm2 $. STIRAP uses a counterintuitive overlapping pulse sequence in which first $L_2$ is pulsed on and then $L_1$. As is well known \cite{Bergmann1998}, STIRAP relies on the existence of a dark state of the form $|D\!\!> = \alpha(t) |a\!\!> + \beta(t) |g\!\!>$. With sufficient adiabaticity, the function $|\alpha(t)|^2$ decreases smoothly from $1$ to $0$, while the function $|\beta(t)|^2$ increases smoothly from $0$ to $1$. The initial state $|a\!\!>$ is thus rotated via $|D\!\!>$ into the final state $|g\!\!>$. The criterion for adiabaticity reads $ \tau_p \Omega^2 \gg (2 \pi)^2 \Gamma $, where $ \tau_p $ is the pulse overlap time, $ \Omega \approx \Omega_1 \approx \Omega_2$ is the peak Rabi frequency during the pulse, and $\Gamma \! \approx \! 2 \pi\times4$ MHz is the (spontaneous) decay rate from the upper state $|e\!\!>$ as determined from our loss measurements. This criterion is quite stringent, in particular in view of the low wave function overlap that enters into $\Omega$, while an upper (experimental) limit for $\tau_p$ is given by the relative laser coherence time for $L_1$ and $L_2$.  We choose $\tau_p $ to be approximately $ 10 \ \mu$s. To get sufficiently high Rabi frequencies given a maximum laser power of up to 26 mW for $L_1$ and $5$ mW for $L_2$ we focus both laser beams to a $1/e^2$-waist of about $25 \ \mu$m. For detection, we apply the reverse STIRAP sequence after a waiting time $\tau_w \!\approx \!10 \ \mu$s to transfer the molecules back into $|a\!\!>$. During this time we leave laser $L_1$ on to assure that all possible residual population in state $|a\!\!>$ is removed.

We perform double STIRAP about $ 3 $ ms after the production of the Feshbach molecules. Fig.~3A shows the molecular population in $|a\!\!>$ as a function of the STIRAP time $\tau$, and Fig.~3B shows the timing sequence for the double transfer scheme. For recording the time evolution of the population we interrupt the transfer process after time $\tau$ and measure the remaining population in $|a\!\!>$. The molecules in $|a\!\!>$ initially disappear during the first STIRAP sequence. They are now in level $|\nu\!=\!73, J\!=\!2\!\!>$ of the singlet $X^1\Sigma_g^+$ potential. Then a large fraction of them returns in the course of the reverse STIRAP sequence. For this particular measurement both lasers are on resonance. The peak Rabi frequencies are $\Omega_1\!\approx\!2\pi\!\times\!3 $ MHz and $\Omega_2\!\approx\!2\pi\!\times\!6$ MHz. We typically obtain an overall efficiency of more than 65\% for the double transfer process, corresponding to single pass efficiencies of more than 80\%, assuming equal efficiencies for both passes. Fig.~3C shows the double pass efficiency as a function of detuning $\Delta_2$ of laser $L_2$. Simulations for the three-level system show that the FWHM-width of the efficiency curve of about $800$ kHz is compatible with a combination of laser power broadening and Fourier broadening. Our simulations also show that higher transfer efficiencies can be expected for an optimized STIRAP pulse sequence in which both peak Rabi frequencies are equal.

We demonstrate coherence of the transfer process in a Ramsey-type experiment \cite{Winkler2007}. For this, we halt the transfer process during the first STIRAP sequence after $\tau \! \approx \! 12 \ \mu$s when a balanced superposition of $|a\!\!>$ and $|g\!\!>$ has been created with $|\alpha(\tau)|^2 \approx \frac{1}{2} \approx |\beta(\tau)|^2 $ by shutting off both lasers instantaneously. After a hold time $\tau_h$ we resume the STIRAP transfer, with the role of lasers $L_1$ and $L_2$ reversed. Thus, for $\tau_h=0$ the population will simply be rotated back into the initial state. A three-level calculation shows that the population in the initial state $|a\!\!>$ is expected to oscillate at the rate of the two-photon detuning $|\Delta_2 - \Delta_1|/(2\pi)$. Fig.~4A shows the initial state population for $\Delta_1\!\approx\!0$ and $\Delta_2\!\approx\!2 \pi\!\times\!113$ kHz as a function of $\tau_h$. Indeed, the population oscillates with a frequency at $|\Delta_2-\Delta_1|/(2\pi)$, however with marked increase in phase jitter on the time scale of $30 \ \mu$s. We attribute this apparent loss of phase coherence to a slow relative frequency drift of lasers $L_1$ and $L_2$, leading to a slightly different two-photon detuning from shot to shot. In Fig.~4A, we have added a region indicating a frequency jitter of $\pm 6$ kHz. This value is compatible with the present stability of our lasers. Note that the frequency drift does not affect an individual STIRAP process as the transfer efficiency is very robust against laser detuning as shown in Fig.~3C.

We now show that the molecular sample is not heated during the transfer process and is indeed in the quantum gas regime. For this we measure and compare the rate of expansion of the molecular sample in state $|a\!\!>$ without and with the double transfer process. In our regime the energy scale for expansion is usually set by the mean field of the BEC, resulting in typical expansion energies for the atoms in the range from $k_B \times 2 $ nK to $k_B \times 10 $ nK, where $k_B$ is Boltzmann's constant, depending on the strength of the atomic interaction \cite{Kraemer2004}. We find that the initial magnetic field ramping excites collective motion of the BEC in the form of a breathing mode as a result of a change in the atomic interaction strength \cite{Weber2003}. The breathing is transformed into expansion of the sample when the trap is shut off. We follow the expansion by monitoring the change of the Thomas-Fermi radius $r$ of the sample. Fig.~4B shows this radius along the horizontal direction as a function of expansion time without and with STIRAP. Without STIRAP, we obtain from a linear fit an expansion rate of $dr/dt\!=\!1.0(1) $ mm/s, corresponding to an energy of $k_B \times 14(4) $ nK. With STIRAP, the rate is $dr/dt\!=\!0.7(1) $ mm/s, corresponding to an energy of $k_B \times 7(2) $ nK. Both values are compatible with the expansion of the atomic sample for the same magnetic field ramp. Interestingly, the rate for the case with STIRAP is lower. We speculate that STIRAP with the tightly focused laser beams $L_1$ and $L_2$ preferentially transfers molecules in the center of the sample and is hence responsible for some selection in velocity space.

It is now possible to add a second STIRAP step for transfer into the rovibrational ground state $\nu\!=\!0, J\!=\!0$. A suitable two-photon transition at readily available laser wavelengths is via the 63rd excited state level of the $ 0_u^+ $ potential near 1339 nm (up) and 999 nm (down) with comparatively good wave function overlap at the level of $|\Omega/\Omega_a|^2 \approx 10^{-4}$ and $10^{-2}$, respectively. We expect that initial spectroscopy near 1339 nm and subsequent search for dark resonances will be straightforward as now all transition energies are known to $10^{-3}$ cm$^{-1}$. Molecules in $\nu\!=\!0, J\!=\!0$ cannot further decay into a lower state upon a two-body collision, and they are thus expected to allow the formation of an intrinsically stable molecular BEC. The high speed of our STIRAP transfer will allow us to perform in-situ as well as time-of-flight imaging for direct characterization of the spatial and momentum distribution of the molecular ensemble. With our technique any low-lying vibrational state can be coherently populated in a controlled fashion with full control over the rotational quantum number, allowing, e.g., state-specific collisional studies and high-precision molecular spectroscopy with possible implications for fundamental physics \cite{Zelevinsky2008,DeMille2008}. Our procedure can be adapted to other species, in particular to heteronuclear
alkali dimers such as RbCs \cite{Sage2005} and KRb \cite{Ospelkaus2008} for the
creation of dipolar quantum gases \cite{Goral2002}. For heteronuclear alkali dimers a
single two-photon transfer step might suffice as a result of favorable wave
function overlap \cite{Stwalley2004}. We expect that the combination of our
technique with Feshbach molecule production out of a Mott-insulator state in a
three-dimensional lattice \cite{Volz2006} will increase the initial Feshbach
molecule production efficiency, avoiding collective excitations as a result of
magnetic field ramping and inhibiting collisional loss, and will provide full
control over all internal and external quantum degrees of freedom of the ground
state molecules.


\bibliographystyle{Science}

\begin{thebibliography}{10}
\bibitem{wavemeter} The wavemeter calibration is currently not sufficient to allow absolute numbering of the frequency comb teeth.
\end{thebibliography}


\begin{scilastnote}
\item We thank the team of J. Hecker Denschlag, the LevT team in our group, and T. Bergeman for
very helpful discussions and M. Prevedelli for technical assistance. We are indebted to
R. Grimm for generous support and gratefully acknowledge funding
by the Austrian Ministry of Science and Research (BMWF) and the
Austrian Science Fund (FWF) in form of a START prize grant and by the European Science Foundation (ESF) in the framework of the EuroQUAM collective research project QuDipMol.
\end{scilastnote}






\clearpage

\begin{center}
\includegraphics[width=12cm]{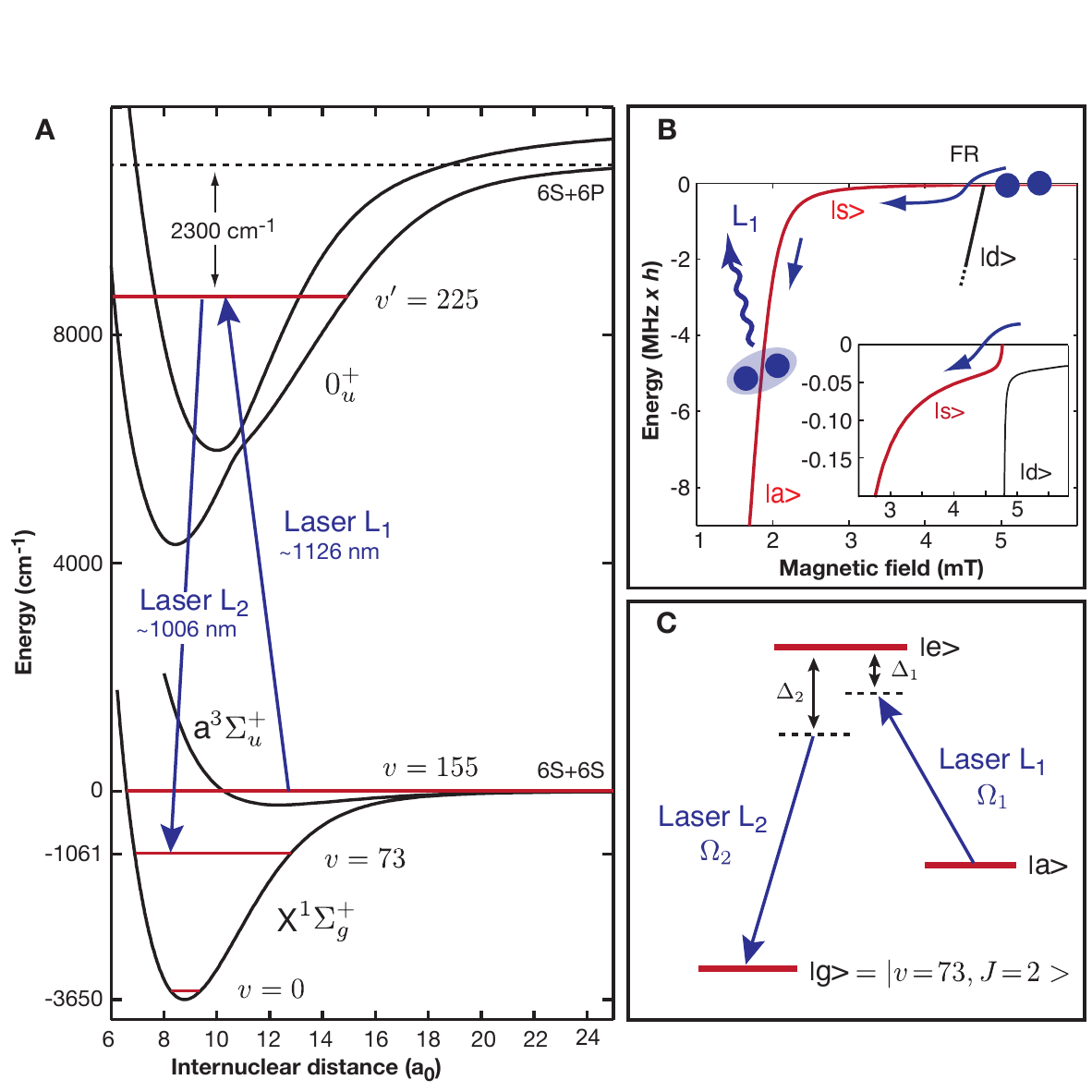}
\end{center}

\noindent {\bf Fig. 1.} {\bf A} Molecular level scheme for Cs$_2$. Molecules in a weakly bound Feshbach level are transferred to rovibrational level $|\nu\!=\!73,J\!=\!2\!\!>$ of the singlet $X^1\Sigma_g^+$ potential with a binding energy of $1061$ cm$^{-1}$ in a two-photon STIRAP process with wavelengths near 1126 nm and 1006 nm via the 225th level of the electronically excited $ (A^1\Sigma_u^+ - b^3\Pi_u) \ 0_u^+$ potentials. The $X^1\Sigma_g^+$ potential has about $155$ vibrational levels. {\bf B} Zeeman diagram showing the energy of all relevant weakly bound molecular levels for initial Feshbach molecular state preparation \cite{Mark2007}. The binding energy is given with respect to the $F\!=\!3, m_F\!=\!3$ two-atom asymptote. The molecules are produced on a d-wave Feshbach resonance at 4.8 mT (see inset) and then transferred to the weakly bound s-wave state $|s\!\!>$ on an avoided state crossing. Further lowering of the magnetic offset field to $1.9$ mT transfers the molecules from $|s\!\!>$ to state $|a\!\!>$, the starting state for the STIRAP transfer. {\bf C} STIRAP transfer scheme \cite{Bergmann1998}. The molecules are transferred from the initial state $|a\!\!>$ to the final state $|g\!\!>=\!|\nu\!=\!73,J\!=\!2\!\!>$ by means of two overlapping laser pulses for which laser $L_2$ is pulsed on first and then $L_1$. The detunings and Rabi frequencies of $ L_i $ are $\Delta_i$ and $\Omega_i$, $i=1,2$.

\clearpage

\clearpage

\begin{center}
\includegraphics[width=12cm]{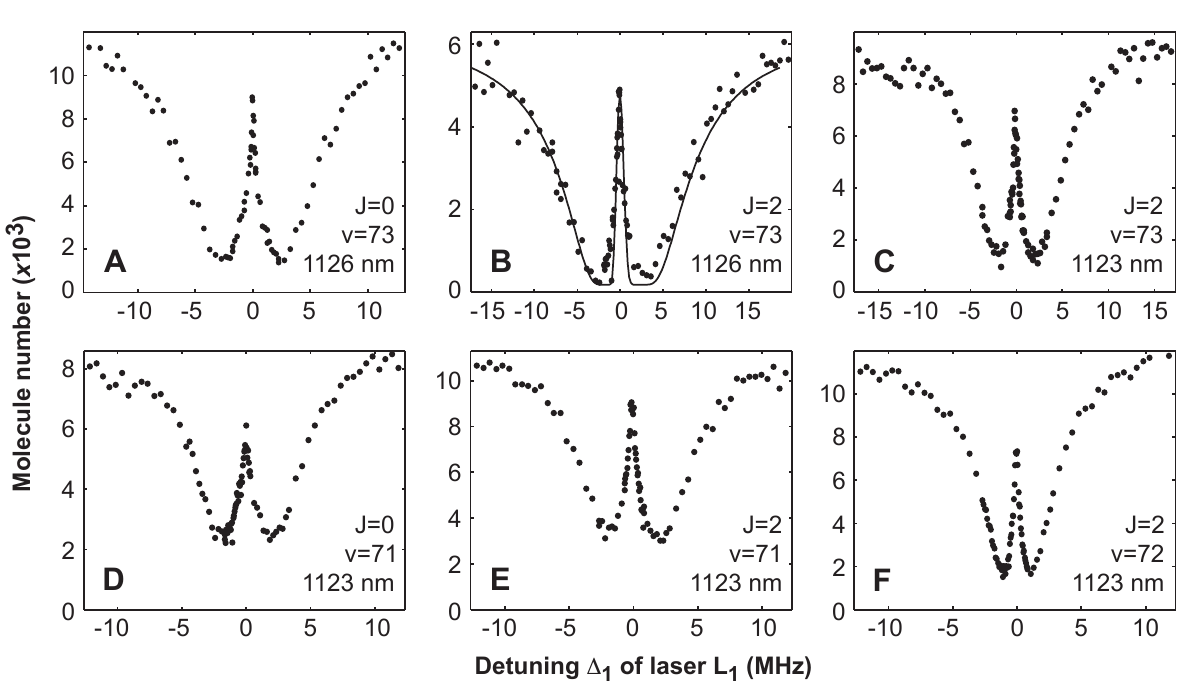}
\end{center}

\noindent {\bf Fig. 2.} Dark resonances for vibrational levels $\nu\!=\!71, 72 $, and $ 73 $. Laser $L_2$ is held on resonance, while the detuning $\Delta_1$ of $L_1$ is scanned. We record the number of molecules in $|a\!\!>$ while both lasers are pulsed on simultaneously. {\bf A}, {\bf B}, and {\bf C} show dark resonances involving  $\nu\!=\!73$ for excitation with $L_1 $ near 1126 nm into $J\!=\!0$ and $2$ and for excitation with $L_1 $ near 1123 nm into $J\!=\!2$, respectively. {\bf D}, {\bf E}, and {\bf F} show the neighboring levels $\nu\!=\!71$ and  $72$ for excitation near 1123 nm. The solid line in B is the result of a three-level model calculation matched to the data giving $\Omega_1 = 2\pi\!\times\!2 $ kHz $\sqrt{I_1 / (\mathrm{mW/cm}^2})$ and $\Omega_2 = 2\pi\!\times\!11 $ kHz $\sqrt{I_2 / (\mathrm{mW/cm}^2}) $ for a pulse time of $5 \ \mu$s at intensities of $ I_1 = 4\!\times\!10^5 $ mW/cm$^2$ for $L_1$ and $ I_2 = 2\!\times\!10^5  $ mW/cm$^2$ for $L_2$ assuming a laser line width of 2 kHz.

\clearpage

\begin{center}
\includegraphics[width=12cm]{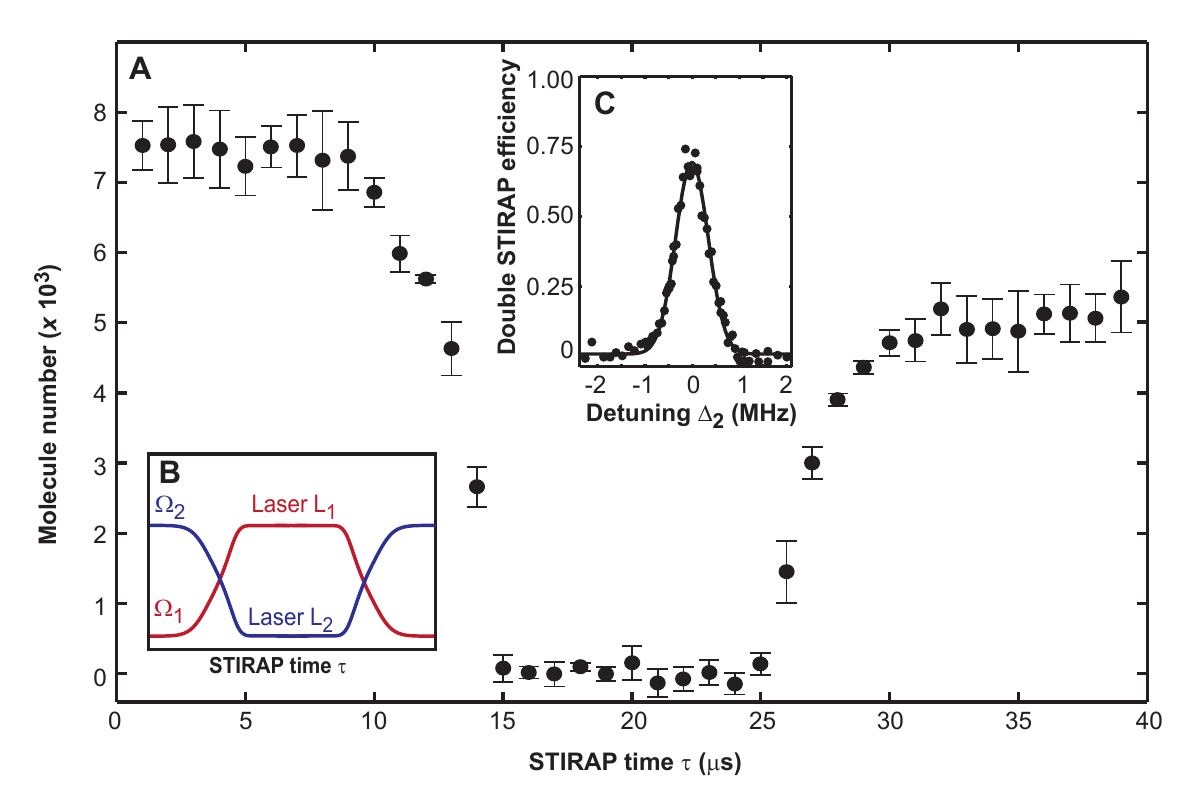}
\end{center}

\noindent {\bf Fig. 3.} STIRAP transfer from the weakly bound state $|a\!\!>$ to the deeply bound state $|g\!\!>=|\nu\!=\!73, J\!=\!2\!\!>$ and back to $|a\!\!>$. {\bf A} shows the number of molecules in state $|a\!\!>$ as a function of STIRAP time $\tau$ for $\Delta_1 \approx 0 \approx \Delta_2$. The measured pulse overlap begins at $5 \ \mu$s and ends at about $15 \ \mu$s. The second pulse overlap starts at $25 \ \mu$s and ends at about $33 \ \mu$s. {\bf B} schematically shows the timing for the Rabi frequencies $\Omega_i$, $i=1,2$, during the double STIRAP sequence. Laser $L_1$ is left on after the first STIRAP sequence to clear out any remaining population in $|a\!\!>$. {\bf C} Double STIRAP efficiency as a function of the detuning $\Delta_2$ of laser $L_2$ for $\Delta_1 \approx 0$. The solid line is a gaussian fit with a FWHM-width of $811$ kHz. The peak Rabi frequencies are $\Omega_1\!\approx\!2\pi\!\times\!3 $ MHz and $\Omega_2\!\approx\!2\pi\!\times\!6$ MHz. The error bars refer to the 1-sigma error in determining the particle number.

\clearpage

\begin{center}
\includegraphics[width=12cm]{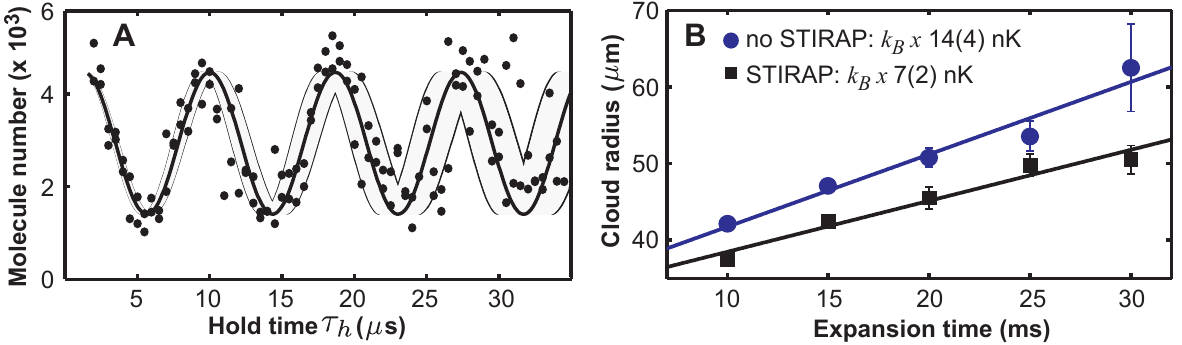}
\end{center}

\noindent {\bf Fig. 4.} {\bf A} Ramsey-type experiment. The population in the initial state $|a\!\!>$ oscillates as the hold time $\tau_h$ during which both transfer lasers are off is increased. The solid line is a sinussoidal fit to the data up to $\tau_h \!=\!20 \ \mu$s. Its frequency $f$ is $115(2)$ kHz. The thin lines are borders to a region that is given by varying $f$ by $\pm 6$ kHz, illustrating the estimated jitter in the two-photon detuning $|\Delta_2 - \Delta_1|$. {\bf B} Comparison of the rate of expansion in the horizontal direction for the molecular sample without and with STIRAP transfer. The top curve (circles) shows the Thomas-Fermi radius $r$ of the molecular sample as a function of expansion time without STIRAP. The linear fit gives a rate of expansion of $ dr/dt = 1.0(1) $ mm/s, corresponding to an energy of $k_B \times 14(4) $ nK. The bottom curve (squares) shows the expansion after double STIRAP with $ dr/dt = 0.7(1) $ mm/s, corresponding to $k_B \times 7(2) $ nK.
\end{document}